\documentclass[%
 aip, jap,
 amsmath,amssymb,
 reprint,%
]{revtex4-1}

\usepackage{graphicx}
\usepackage{dcolumn}
\usepackage{bm}

\usepackage[utf8]{inputenc}
\usepackage[T1]{fontenc}
\usepackage{mathptmx}
\usepackage{etoolbox}

\usepackage{xcolor}
\usepackage{soul}

\usepackage[unicode]{hyperref}

\hypersetup{colorlinks=true,
    linkcolor=purple,
    citecolor=purple, urlcolor=purple, breaklinks=true}

\usepackage{multirow}
\usepackage{colortbl}
\usepackage{booktabs}
\usepackage{longtable}
\usepackage{makecell,tabularx}
\usepackage{layout}

\makeatletter
\def\@email#1#2{%
 \endgroup
 \patchcmd{\titleblock@produce}
  {\frontmatter@RRAPformat}
  {\frontmatter@RRAPformat{\produce@RRAP{*#1\href{mailto:#2}{#2}}}\frontmatter@RRAPformat}
  {}{}
}%
\makeatother
\begin{document}

\preprint{AIP/123-QED}

\title[Epitaxial rare-earth pyrochlore iridates]{Epitaxial thin films of pyrochlore iridates: a forward looking approach}
\author{Araceli Gutiérrez-Llorente}
\email{araceli.gutierrez@urjc.es}
\affiliation{
Escuela Superior de Ciencias Experimentales y Tecnología, Universidad Rey Juan Carlos, Madrid 28933, Spain
}%



\begin{abstract}

Topological quantum materials that show strongly correlated electrons as well as topological order, for which spin-orbit coupling is a key ingredient, exhibit novel states of matter. One such example is the family of pyrochlore iridates, featuring strong spin-orbital coupling, strong electron interactions as well as geometric frustration, making them an ideal platform to study novel topological phases. High-quality epitaxial pyrochlore iridate films, although challenging to produce, provide a pathway to explore unconventional behaviours and unravel the intrinsic properties of these largely unexplored materials. Additionally, designing interfaces with specific properties is crucial to create multilayered devices that can achieve significant technological breakthroughs using topological states of these materials.  This article reviews experimental work on epitaxial pyrochlore iridate thin films,  discussing evidence of topological phases found in them. Future research directions are outlined, which include exploring the rich tunability offered by chemical doping, especially when combined with the design of epitaxial heterostructures.

\end{abstract}

\maketitle

\section{\label{sec:intro}Introduction}

Crystalline transition metal oxides (TMO), characterized by strong electron interactions within $d$-orbitals, are a source of rich physics.  These materials can exhibit nearly every known electronic phase.\cite{imada:98,tokura:00,dagotto:05,georges:13,ngai:14}  Moreover, remarkable emergent phenomena occur at the interface between different oxides,\cite{ohtomo:04,reyren:07,mannhart:08,caviglia:10,jang:11,zubko:11,hwang:12,stemmer:14,yadav:16,huang:18,ramesh:19} underscoring their significant potential for integration into electronic devices.\cite{yang:11,coll:19,vaz:21,park:23}

Up to now, extensive research has focused on the broad family of oxides crystallizing in the cubic ABO$_3$ perovskite structure.\cite{goodenough:04}  Extending this field of research to oxides with alternative crystal structures remains at the forefront of quantum materials research,\cite{keimer:17,tokura:17,samarth:17,basov:17,giustino:20,ahn:21,sobota:21,paschen:21,checkelsky:24,wang:24} opening up an opportunity to reveal exciting and unconventional behaviours in condensed matter physics.

Likewise oxides with perovskite structure, A$_2$B$_2$O$_7$ oxides that adopt the pyrochlore structure\cite{subramanian:83} display a wealth of physical properties.  These properties include magnetism with exotic ground states (e.g. spin ice),\cite{gardner:10,bramwell:98,bramwell:01,greedan:06} ferroelectricity,\cite{nunez-valdez:19} piezoelectricity,\cite{ishibashi:10} superconductivity,\cite{hanawa:01,sakai:01,hiroi:18} colossal magnetoresistance,\cite{subramanian:96,ramirez:97} catalytic activity,\cite{oh:12,kim:17,lebedev:17,kim:20a} ionic conductivity,\cite{fuentes:24,gayen:22,supriya:23,lian:01} and high tolerance to radiation damage.\cite{ewing:04,sickafus:00}  This diversity stems partly from the extensive range of possible substitutions at both the A and B sites, akin to perovskites.  Yet, this is not the only factor at play.

The pyrochlore crystal structure contains two interpenetrating pyrochlore lattices formed of corner-sharing tetrahedra, each occupied by a different species of cation.  Either or both sublattices can host magnetic moments, but the magnetic interactions are frustrated from the geometry, hence failing to develop long-range magnetic order.  This geometry is at the origin of a wealth of fascinating phenomena, such as the unconventional anomalous Hall effect driven by spin chirality even without long-range magnetic order of conduction electrons,\cite{machida:07,machida:10,ueda:12,fukuda:22} and fragmentation of the magnetic moment field, in which spin fluctuations coexist with an ordered phase.\cite{brooks-bartlett:14,henley:10,petit:16,lefrancois:17,lhotel:20} As a result of the strong frustration, the pyrochlore lattice also hosts fractional quasiparticles (emergent excitations of collective behaviour), for example, emergent magnetic monopoles arising from fractionalization of magnetic dipoles in spin ice.\cite{ryzhkin:05,castelnovo:08,bramwell:20,fennell:09,bramwell:09,jaubert:09,morris:09,kadowaki:09} 

In recent years, the the study of the topology of the electronic structure of crystalline materials has become increasingly central in the research on strongly correlated electron materials.  This shift in perspective transcends traditional band theory of crystalline solids, embracing the topological properties (topological invariants) that these energy bands can exhibit,\cite{bansil:16,narang:21,cayssol:21,wieder:22} thereby transforming our understanding of matter.

Spin-orbit interaction plays a central role in the emergence of topological phases,\cite{jackeli:09,pesin:10,witczak:14,schaffer:16}  intrinsically different from the conventional phases characterized by long-range order and a change in some symmetry.  Notably, in 5$d$-transition metal oxides, such as iridium oxides or iridates, the strength of electronic correlation and spin-orbit coupling is of the same order.  This makes them ideal candidates for the emergence of topological phenomena, entirely different from those observed in 3d-transition metal oxides.\cite{witczak:12,rau:16,cao:18,takayama:21}

Within this context, the rare-earth pyrochlore iridates R$_2$Ir$_2$O$_7$ (R= Y or a lanthanide element) provide a unique opportunity to study the interplay of fundamental interactions, in which Coulomb repulsion among electrons can be tuned by changing the ionic radius of the R cation,\cite{matsuhira:07,yang:10,chen:12,savary:14,yang:14,ueda:16,yamaji:16,goswami:17,ueda:17,oh:18} and band filling of the Ir 5d state can be controlled through doping.\cite{kaneko:19,ueda:20}  Except for Pr$_2$Ir$_2$O$_7$, which remains metallic down to the lowest temperature, pyrochlore iridates  feature non collinear All–In–All–Out (AIAO) magnetic order for the Ir-5$d$ moment below the metal-insulator transition (MIT), whose temperature decreases with increasing the $R$ ionic radius.\cite{matsuhira:07}  No evidence of any crystal symmetry change has been found associated with this transition.\cite{shapiro:12,clancy:16,takatsu:14}

All this gives rise to an interesting scenario. For instance, the topological Weyl semimetal phase was first predicted in the pyrochlore iridates.\cite{wan:11,yang:11,witczak:12,nagaosa:20,lv:21} Although it has not been experimentally achieved yet, experimental signatures of field-induced emergent states, which may be potentially correlated to the predicted topological states were revealed, such anomalous Hall effect,\cite{machida:07,ueda:18} highly conductive magnetic domain walls in a magnetic insulator,\cite{ueda:14,ma:15} and magnetic field induced MIT.\cite{ueda:15a,tian:16,ueda:17}

To deeply investigate the fascinating phenomena emerging in these materials and experimentally verify the theoretically predicted properties, it is crucial to grow them as epitaxial films with precise control over crystal growth orientation.  This demand arises because magnetic order is highly anisotropic,\cite{hozoi:14} requiring field-directional studies on single crystals. Moreover, confined geometries result in film properties that are markedly different from those in bulk.\cite{yang:14,schlom:08}  Additionally, high-quality epitaxial films are also essential for heterostructure engineering and, ultimately, for the development of devices.  However, growth of epitaxial pyrochlore iridate films poses a unique set of challenges.\cite{evans:18,kim:22,hensling:24}

These challenges encompass the difficulty of achieving stoichiometric compositions in the films due to the high volatility of iridium in oxygen (further details will be provided later),\cite{alcock:60,nair:23} and managing disorder that can significantly impact ground states, hypothetically even fostering new disordered phases in some cases.\cite{savary:17a}  Hence a strong correlation between the physical properties  and the quality of epitaxial pyrochlore oxide films is anticipated, potentially leading to sample-dependent low-temperature experimental outcomes.  This interconnection is a hallmark of systems with intricate magneto-structural properties and large ground-state degeneracies, in which the magnetic field (or other external perturbation) can favour one of various competing phases harboured by the material,\cite{taniguchi:13,jaubert:15,bowman:19} yet the specific mechanisms remain puzzling.

Furthermore, the scarcity of pyrochlore substrates poses a significant obstacle for experimental studies. This limitation hinders progress, since the ideal substrate share the same structure as the film being grown, thereby determining the epitaxial strain, which greatly impacts the physical properties of the films, as demonstrated within the structural family of perovskites.\cite{haeni:04,lee:10,schlom:14}.  The wide variety of available single-crystal substrates with cubic and pseudocubic perovskite structures\cite{uecker:17} allows for the tailoring of physical properties in epitaxial perovskite films by tuning strain.\cite{sando:22,li:24}. In sharp contrast, pyrochlore substrates have only recently become commercially available and remain rare.  Consequently, yttria-stabilized ZrO$_2$ (YSZ) substrate with fluorite structure is the preferred choice for epitaxially growing pyrochlore films, as the pyrochlore structure can be viewed as an ordered defect fluorite with a lattice parameter approximately twice that of YSZ. Nevertheless, this alternative complicates strain engineering in pyrochlore films,\cite{ohtsuki:19,guo:20,kim:20b} making it challenging to explore the tuning of their electronic state topology through lattice strain.

\begin{figure*}[htb]
 \includegraphics[keepaspectratio=true, width=1.0\linewidth]{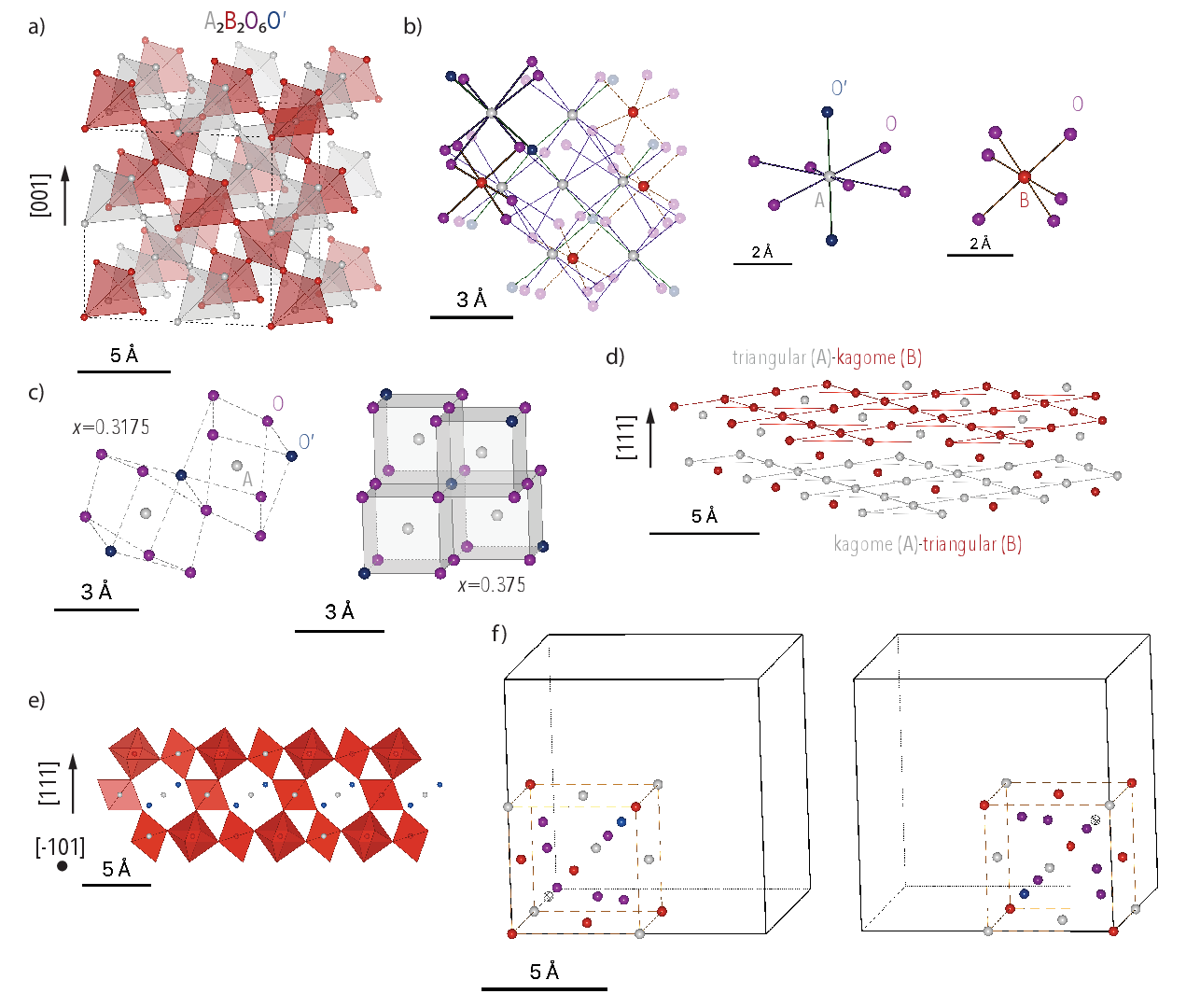} \caption{\label{Fig_struct_01}  {{\bf{Pyrochlore structure.}}  {\bf{(a)}} The pyrochlore lattice A$_2$B$_2$O$_6$O$^{\prime}$ features two distinct sublattices of A and B sites, with oxygen anions at the tetrahedra vertices are omitted for clarity.  {\bf{(b)}} The coordination environment of A and B ions: A ions exhibit eight-fold coordination, while B ions show six-fold coordination. A cations are bonded to six O atoms and two O$^{\prime}$ atoms, with the A-O$^{\prime}$ bond being shorter than the A-O bond. The  O$^{\prime}$-A-O$^{\prime}$ axis is perpendicular to the mean plane of the six 0 atoms.  {\bf{(c)}} Coordination around A ions varies with the oxygen position paramenter $x$.  {\bf{(d)}} Along the [111] direction, the pyrochlore lattice alternates between kagome and triangular planar layers.  {\bf{(e)}} View along $[\bar{1}01]$: B ions are in octahedral coordination (corner-linked BO$_6$ units, with the oxygen atoms at the vertices omitted for clarity) and A$_2$O$^{\prime}$ chains fill the interstices.  {\bf{(f)}} The arrangement of A$^{3+}$ and B$^{4+}$ along $\langle110\rangle$ results in two types of fluorite subcells within the pyrochlore structure.
}}
\end{figure*}

This short review aims to provide a comprehensive overview of experimental studies on epitaxial pyrochlore iridate films, addressing the challenges inherent to this class of materials and briefly discussing future directions for advancing experimental research. Studies on bulk single crystals or polycrystalline films of pyrochlore iridates are excluded in this article, except as references for materials properties. While the primary focus of this review is on experimental work, it also seeks to bridge the gap between tantalizing theoretical predictions and their experimental validation in thin films. 

Although this review has limited scope, it is not possible to provide complete reference list. This limitation also applies to the growing number of theoretical studies on topological phases of matter that may be realized in these materials.  Readers are referred to several excellent review articles that delve into various related aspects of the subject. Notably, reviews of emergent quantum phases arising from the interplay of strong electron interactions and spin-orbit coupling, illustrated by pyrochlore iridates, were given by Pesin and Balents,\cite{pesin:10} Witczak-Krempa \textit{et al.},\cite{witczak:14} Savary \textit{et al.},\cite{savary:14} Schaffer \textit{et al.},\cite{schaffer:16} Goswami \textit{et al.},\cite{goswami:17} and Cao and Schlottmann.\cite{cao:18} Moreover, the work by Chakhalian \textit{et al.}\cite{chakhalian:20} emphasizes the rich possibilities of TMO thin films and heterostructures grown along the [111] direction of perovskite and pyrochlore lattices. 

In this paper current state of the field of epitaxial pyrochlore iridate thin films is reviewed.  In Section II, the geometrically frustrated lattice structure of pyrochlores is described, since it plays a crucial role in the emerging electromagnetic functions of pyrochlore iridates. Next, section III reviews the magnetic structure of pyrochlore iridates. Section IV discusses research on epitaxial pyrochlore iridate films, and is divided into two subsections about epitaxial growth, in which a survey of deposition methods and specific challenges is presented, and experimental evidence of physical properties of films with emphasis on strain effects.  The review closes with an outlook on future directions of research in this field, including chemical doping, and design and growth of epitaxial heterostructures.

\section{\label{sec:pyrochlore structure}Pyrochlore structure}

The structure of pyrochlore oxides A$_2$B$_2$O$_6$O$^{\prime}$ (usually written as A$_2$B$_2$O$_7$, cubic $Fd\bar{3}m$, space group 227) is a geometrically frustrated lattice with four non-equivalent crystallographic positions.\cite{subramanian:83,gardner:10} This structure comprises two distinct sublattices of A and B sites, which are structurally identical but displaced by half a lattice constant from each other, Fig.\ref{Fig_struct_01}(a).  The ions A and B are located at the vertices of corner-sharing tetrahedra in each sublattice.  In the compounds discussed in this paper, A$^{3+}$ represents a rare earth cation, while B$^{4+}$ denotes the tetravalent cation of iridium (Ir$^{4+}$). The oxygen anions sit at two distinct crystallographic sites, identified as O and O$^{\prime}$.  The A cations show an eight-fold coordination geometry, with six A-O bonds and two shorter A-O$^{\prime}$ bonds. The B cations are six-coordinated with oxygen anions, positioned within trigonal antiprisms (BO$_6$), Fig.\ref{Fig_struct_01}(b).

The oxygen position parameter $x$ ($0.3125\leq x \leq 0.375$) controls the position of the O atoms and consequently the coordination geometry around the cations  This parameter describes the trigonal compression of BO$_6$ octahedra, resulting in regular octahedra around B cations for $x=0.3125$. For $x=0.375$ regular cubes form around A cations, which corresponds to a fully disordered material akin to a fluorite lattice, Fig.\ref{Fig_struct_01}(c).  This parameter is highly sensitive to structural disorder.\cite{minervini:02}

Along the [111] direction the pyrochlore lattice comprises alternating kagome (the most frustrated 2D geometry for magnetism) and triangular planar layers, which serve as natural cleave planes, Fig.\ref{Fig_struct_01}(d). Consequently, this direction is ideal for film growth in pyrochlores and is expected to promote the formation of topological phases in oxide thin films.\cite{hu:12,yang:14,chen:15}  Additionally, along the direction $[\bar{1}01]$ the pyrochlore structure consists of corner-shared BO$_6$ distorted octahedra with the A cations occupying the interstices, Fig.\ref{Fig_struct_01}(e).

\begin{figure}[htb]
 \includegraphics[keepaspectratio=true, width=1.0\linewidth]{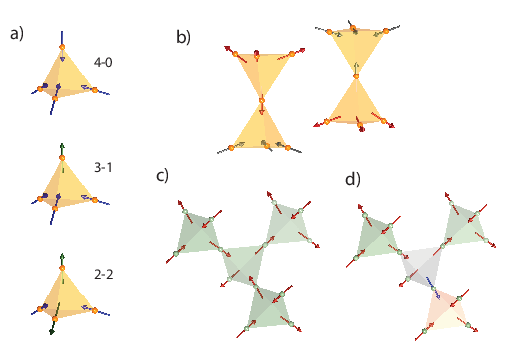} \caption{\label{Fig_magnetic_moments}  {{\bf{Magnetic ions in the pyrochlore lattice.}} {\bf{(a)}} Different arrangements of spins in a single tetrahedron with $\langle111\rangle$ Ising anisotropy.  Spins align along the local direction from a vertex to the center of the tetrahedron in a configuration that can be 4-0, with the four spins pointing in; 3-in 1-out; or, 2-in 2- out.  {\bf{(b)}} Two distinct magnetic domains All-in-All-out/All-out-All-in of the 4-0 configuration.  {\bf{(c)}} Spin ice state in a pyrochlore lattice in which each tetrahedron has the 2-2 configuration.  In this case, a single spin flip {\bf{(d)}} produces elementary excitations on two neighbouring tetrahedra that can be regarded as fractionalized magnetic monopoles.}}
\end{figure}

The arrangement of A$^{3+}$ and B$^{4+}$ ions along the direction [110] defines two types of fluorite subcells within the pyrochlore lattice, Fig.\ref{Fig_struct_01}(f).  Thus, the pyrochlore structure can also be described as a superstructure composed of eight nearly identical fluorite subunits, with intrinsic ordered oxygen vacancies at 1/8 of the oxygen sites to maintain charge neutrality.

The pyrochlore lattice exhibits a strong tendency towards disorder at both cation and oxygen sites.  This disorder can be induced by substitution of A cations by B ones, with simultaneous rearrangements in both sublattices, or by the randomisation of oxygen vacancies, through the migration of oxygen to the nearest vacant site.\cite{minervini:00,minervini:02,talanov:21}  Eventually, the pyrochlore can transform into a disordered fluorite structure by mixing the A and B ions within the cation sublattices and randomly distributing the oxygen vacancies.  This predisposition to disorder is more pronounced in compounds with small A cations and large B cations.\cite{minervini:00}  Additionally, lattice displacements and local off-centering of the A and O ions were also reported.\cite{koohpayeh:14,trump:18}

It is well established that weak disorder can lead to spin-glassy behaviour in pyrochlore magnetic oxides.\cite{booth:00,andreanov:10}  Interestingly, rather than yielding glassy behaviour, it was suggested that weak disorder introduced by structural imperfections around the magnetic sites in non-Kramers ions (such as Pr$^{3+}$ or Tb$^{3+}$) of pyrochlore magnetic oxides can give rise to a quantum spin liquid ground state.\cite{savary:17a} To test this hypothesis, experimental studies\cite{wen:17,martin:17} on Pr$_2$Zr$_2$O$_7$ focused on diffuse neutron scattering experiments around Bragg peaks caused by structural distortions.  These experiments revealed a level of disorder undetectable by conventional diffraction, but sufficient to induce quantum fluctuations in spin.  However, a quantum spin liquid state induced by disorder has yet to be experimentally confirmed.\cite{benton:18}  In this context, it is noteworthy that local atomic arrangements in disordered crystalline materials can be nonrandom, giving rise to short-range order at a length scale shorter than the unit cell.\cite{oquinn:20} The impact of this short-range order on physical properties remains to be fully understood.

\section{\label{magnetic order} Magnetic order/ structure in pyrochlore iridates}

Except for Pr$_2$Ir$_2$O$_7$, which remains magnetically disordered in a metallic state down to the lowest temperatures,\cite{nakatsuji:06,tokiwa:14} pyrochlore iridates exhibit antiferromagnetism below the metal-to-insulator transition.  This  transition is characterized by a non collinear All–In–All–Out (AIAO)  ordering of the Ir$^{4+}$ sublattice in the magnetic insulating phase.  The AIAO configuration can transform into a 2-2 or 3-1 structure when a magnetic field is applied along specific crystal axis, significantly affecting the magnetic and electronic properties of the material, Fig.\ref{Fig_magnetic_moments}.  This non-coplanar spin arrangement is at the origin of the spin-chirality induced anomalous Hall effect.\cite{taguchi:01}  Moreover, nontrivial topological electronic bands, closely related to this magnetic order, have been predicted.\cite{witczak:12,go:12,witczak:13,imada:14,ueda:16,goswami:17,ladovrechis:21}

There is experimental evidence of the AIAO magnetic arrangement in the iridium sublattice in Nd$_2$Ir$_2$O$_7$,\cite{guo:16,asih:17,tian:16,ueda:15b} Eu$_2$Ir$_2$O$_7$,\cite{sagayama:13,fujita:15,fujita:16a}  Tb$_2$Ir$_2$O$_7$,\cite{lefrancois:15,fujita:16b,donnerer:19} Er$_2$Ir$_2$O$_7$,\cite{lefrancois:15} Sm$_2$Ir$_2$O$_7$;\cite{donnerer:16,asih:17} Yb$_2$Ir$_2$O$_7$ and Lu$_2$Ir$_2$O$_7$.\cite{jacobsen:20}

Although the magnetic domains are antiferromagnetic, their alignment can be controlled through a magnetic field cooling procedure. And the magnetic domain structure can be detected through magnetotransport measurements, as spin-orbit coupling renders conduction electrons responsive to the local spin arrangement.  For example, magnetic-field induced AIAO/AOAI domain inversion has been reported in epitaxial films of Eu$_2$Ir$_2$O$_7$ depending on the polarity of the cooling magnetic field in the [111] crystallographic direction\cite{fujita:15,arima:13} (see section \ref{film_properties}).

\begin{figure}[htb]
 \includegraphics[keepaspectratio=true, width=0.8\linewidth]{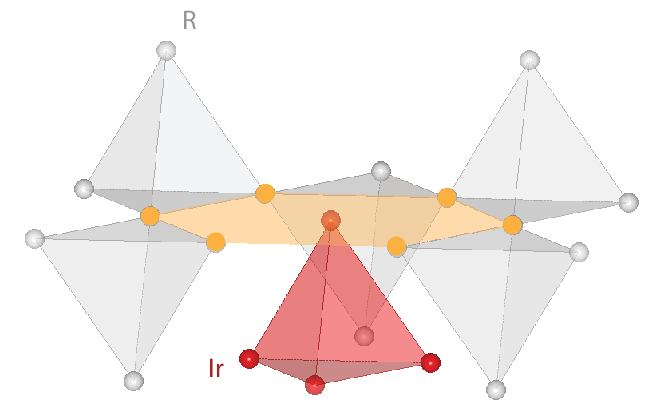} \caption{\label{Fig_exchange_f_d}  {{\bf{R 4$f$—Ir 5$d$ exchange coupling in R$_2$Ir$_2$O$_7$.}}  If the ion R$^{3+}$ carry a magnetic moment, the application of an external magnetic field can control the R spin configuration which, in turn, modulates the Ir magnetic ordering. A single Ir$^{4+}$ ion (red) is surrounded by six nearest-neighbouring R-4 sites (yellow) whose moments generate an effective magnetic field at the Ir site. Moreover, below the metal-insulator transition temperature, when the Ir$^{4+}$ sublattice orders in the AIAO phase, a magnetic field oriented along the local [111] directions on the R$^{3+}$ sublattice is generated.
}}
\end{figure}

The rare-earth cation sublattice can also be magnetic in rare-earth pyrochlores, leading to even more complex magnetic structures. The \textit{f}-electrons of the rare-earth R$^{3+}$ ion can carry a net magnetic moment in a Kramers doublet (R = Nd, Sm, Gd, Dy, Yb) or a non-Kramers doublet (Pr, Tb, Ho).  The $f$-$d$ exchange interaction between the localized R-4$f$ electron moments and the more itinerant Ir-5$d$ electrons results in the modulation of the magnetic order of Ir$^{4+}$ moments by the molecular fields of R$^{3+}$ magnetic ions (Fig.\ref{Fig_exchange_f_d}). Theoretical studies suggested that this exchange interaction between Ir and R ions is crucial for stabilizing topological phases in pyrochlore iridates, such as the Weyl semimetal and the axion insulator.\cite{chen:12}

For example, in polycrystalline samples of Tb$_2$Ir$_2$O$_7$, the Ir molecular field induces the ordering of Tb magnetic moments below 40 K into an AIAO magnetic arrangement, while  no evidence of magnetic long-range order was observed in the Er sublattice of Er$_2$Ir$_2$O$_7$ down to 0.6 K.\cite{lefrancois:15}  This different behaviour was attributed to the anisotropy of the R site. The molecular fields from Ir$^{4+}$ in the AIAO order induce the Tb$^{3+}$ moments to adopt a AIAO configuration at low temperature due to their strong axial magnetocrystalline anisotropy along the ⟨111⟩ direction. In contrast, Er exhibits easy-plane anisotropy, with its magnetic moments perpendicularly oriented to the Ir molecular field.

Nd-4$f$ moments also develop the AIAO-type ordering at lower temperature due
to $f$-$d$ exchange coupling in Nd$_2$Ir$_2$O$_7$.\cite{tomiyasu:12,ueda:15b} The onset temperature of the AIAO-type magnetic ordering of the Ir 5$d$ moment is 15 K, and domain flipping between the 4-0 and 0-4 configuration is driven by a magnetic field along [111].  At 2 K, a magnetic field $H \parallel [111]$ modulated by $f$-$d$ exchange coupling can switch between the two variants AIAO/AOAI in the Ir 5$d$ configuration and from 0-4 to 3-1 in the Nd 4$f$ configuration as $H$ increases. In contrast, both Nd and Ir sublattices develop a 2-2 out configuration under a magnetic field $H \parallel [001]$.\cite{ueda:15b} 

Gd$_2$Ir$_2$O$_7$ and Ho$_2$Ir$_2$O$_7$ exhibit complex magnetic order patterns.  The ordering temperature of the Ir$^{4+}$ sublattice in an AIAO structure, which coincides with the metal-to-insulator transition, is approximatively 127 K and 140 K for polycrystalline samples of Gd$_2$Ir$_2$O$_7$ and Ho$_2$Ir$_2$O$_7$, respectively.  The ordering temperatures for Gd$^{3+}$ and Ho$^{3+}$ are around 50 K (induced by the AIAO Ir molecular field through Ir—Gd magnetic coupling) and 18 K, respectively.\cite{lefrancois:19,lefrancois:17}

Gd$_2$Ir$_2$O$_7$ and Ho$_2$Ir$_2$O$_7$ display different types of anisotropy.  Gd$^{3+}$ in Gd$_2$Ir$_2$O$_7$ does not exhibit strong single-ion anisotropy, although a weak easy-plane anisotropy for Gd$^{3+}$ emerges below 2 K, owing to the mixing of its ground state with higher spectral multiplets, that can result in anisotropic exchange.\cite{lefrancois:19}  Ho$^{3+}$ in Ho$_2$Ir$_2$O$_7$ shows an easy-axis anisotropy along the [111] direction and fragmentation of the magnetisation below 2 K into two distinct components:\cite{lefrancois:17} an ordered antiferromagnetic phase and a disordered Coulomb phase with ferromagnetic correlations.\cite{brooks-bartlett:14}  Quantitative analysis of the interplay between the molecular fields of the AIAO ordered Ir sublattice and an external magnetic field on experimental data from polycrystalline samples is challenging, but Monte Carlo simulations carried out to elucidate the underlying physics of the unconventional magnetic ground state indicate that applying an external magnetic field applied along different directions can induce the stabilization of exotic phases, such as fragmented kagome ice state, characterize by fragmented phases in the kagome planes and spins aligned along the magnetic field in the triangular planes.  On the other hand, experimental measurements on single crystals of Ho$_2$Ir$_2$O$_7$ combined with Monte Carlo simulations indicate that an applied magnetic field along the direction [111] arranges the Ho moments into a 3-1 configuration, favouring one type of configuration of AIAO domains in Ir (conversely, applying the field in the opposite direction favours the alternate configuration in the Ir sublattice). In contrast, applying the magnetic field along [100] orders the Ho spins into a 2-2 state.\cite{pearce:22}

Fragmentation of magnetic moments between disordered and ordered components was also observed in Dy$_2$Ir$_2$O$_7$. Experimental measurements of polycrystalline and single crystal samples of Dy$_2$Ir$_2$O$_7$ show that it stabilizes the fragmented monopole crystal state, in which antiferromagnetic AIAO order coexists with a Coulomb phase spin liquid, at temperatures below around 1 K.\cite{cathelin:20}

Geometrically frustrated magnets are known to have strong spin-phonon coupling.\cite{tchernyshyov:11,castellano:15,lummen:08} In 5d materials, such as the pyrochlore iridates, the Dzyaloshinskii-Moriya (DM) interaction, which originates from spin-orbit coupling, may induce spin-phonon coupling.\cite{son:19}

Strong spin-phonon coupling has been reported in polycrystalline samples of the pyrochlore iridate Y$_2$Ir$_2$O$_7$ by infrared spectroscopy and first-principles calculations.\cite{son:19} Some Ir-active phonons soften and sharpen below the antiferromagnetic ordering temperature at 170 K. Since the AIAO magnetic ordering in pyrochlore iridates is not accompanied by structural transitions, the observed phonon anomalies must be due to the coupling to the AIAO magnetic order.  The authors attributed these anomalies to the modulation of the Ir—O—Ir bond angle by the DM interaction.

Furthermore, Ueda {\textit{et al.}}\cite{ueda:19} studied Eu$_2$Ir$_2$O$_7$ using Raman scattering. Their research unveiled large phonon anomalies triggered by the emergence of magnetic AIAO order in Eu$_2$Ir$_2$O$_7$, below 115 K. Notably, Ir—O—Ir bond bending vibration exhibited line-shape anomalies at the magnetic phase transition, indicating strong electron-phonon interactions that are crucial for understanding the magnetic ordering behaviours of pyrochlore iridates. The authors proposed a spin-phonon coupling mechanism modulated by exchange interactions, widely accepted for 3d materials; phonon-induced modulation of the orbitals, which affects magnetism via spin-orbit coupling; and electron-phonon coupling through charge fluctuations. However, the microscopic mechanism responsible for the electron-phonon interactions that the experimental data reveal remains unresolved.

Coupling of phonons to spins has also been studied in Pr$_2$Ir$_2$O$_7$ by thermal transport measurements in single crystals.\cite{uehara:22} A strong resonant scattering between phonons and paramagnetic spins was found, hindering longitudinal heat conduction.

\section{\label{sec:epitaxial_films}Epitaxial pyrochlore iridate films}

\setlength{\tabcolsep}{7pt}


\begin{table*}
\caption{\label{tab:table1}Single-crystal pyrochlore iridate films synthesized up to date.}
\renewcommand{\arraystretch}{1.3}
\begin{tabular}{p{3cm}p{13.5cm}}
\toprule
\thead{Material | Substrate} &  \thead{Growth process} \\\midrule
\multirow{3}{*}{Pr$_2$Ir$_2$O$_7$ | YSZ(111)} & $\bullet$ SPE:\footnote{Solid-phase epitaxy.} PLD (Ir-rich target Pr:Ir=1:2, RT,\footnote{RT: Room temperature.} 10 mTorr, 0.65 Jcm$^{-2}$, 5 Hz), post-annealing at 1000$^{\circ}$C, 1.5 h in air.\cite{ohtsuki:19,ohtsuki:20} 
Conventional PLD at 10 mTorr, 0.65 Jcm$^{-2}$, 5 Hz, T up to 800$^{\circ}$C, using stoichiometric and Ir-rich (Pr:Ir=1:2) targets, results in insulating films, consistent with Pr$_2$O$_3$ phase instead of metallic Pr$_2$Ir$_2$O$_7$.\cite{ohtsuki:20}  \\

    &$\bullet$ SPE: RF sputtering (polycrystalline Pr$_2$Ir$_2$O$_7$ target, RT, 15 mTorr in Ar:O$_2$=90:10), post-annealing at 800$^{\circ}$C, 12 h in air.\cite{guo:20}  \\

    &$\bullet$ Partially strained films. First step: RRHSE.\footnote{RRHSE: Repeated rapid high-temperature synthesis epitaxy. See also Fig. \ref{Fig_2023_Song_fig_1}.} Alternate Pr$_2$Ir$_2$O$_7$ and IrO$_2$ amorphous layers in each cycle by PLD at 600$^{\circ}$C, 50 mTorr, 1.5 Jcm$^{-2}$, 5 Hz. Then the temperature is increased to 800$^{\circ}$C for 30 s, and rapidly decreased to 600$^{\circ}$C for the next cycle. Second step: post-annealing at 1000$^{\circ}$C in a sealed tube with IrO$_2$ powder for 1 h.\cite{li:21}\\

    &$\bullet$ Experimental and computational study of \textit{in situ} phase formation in the Pr-Ir-O$_2$ system. Co-sputtering deposition: the flux of Pr$_2$Ir$_2$O$_7$ and IrO$_2$ is controlled separately.\cite{guo:21}  Pr$_2$Ir$_2$O$_7$ can only form high-quality crystals with $T>1073$ K and $P_{O_{2}}>9$ Torr.\\\cmidrule[0.5pt]{1-2}

\multirow{4}{*}{Nd$_2$Ir$_2$O$_7$ | YSZ(111)} &$\bullet$ SPE: RF sputtering (RT, 12.5 mTorr in Ar:O$_2$=99:1), post-annealing at 750$^{\circ}$C,  for 12 h, with oxygen partial pressure between 30 mbar and 200 bar. Annealing at $\geq 775^{\circ}$C results in the formation of impurity phases.\cite{gallagher:16}\\ 
&$\bullet$ SPE: PLD (600$^{\circ}$C, 50 mTorr, 4.5 Jcm$^{-2}$ and 3 Hz), post-annealing at 1000$^{\circ}$C, 1 h in air.\cite{kim:18}\\
&$\bullet$ Strained films.\cite{kim:20b,song:23,shen:24} RRHSE. Stoichiometric amorphous Nd$_2$Ir$_2$O$_7$ and IrO$_2$ deposited by PLD at T  600$^{\circ}$C, 50 mTorr, 5 Hz and 4.5 Jcm$^{-2}$(ref.\cite{kim:20b}) or 1.7 Jcm$^{-2}$(ref.\cite{song:23}). Then, rapid thermal annealing, T up to 800$^{\circ}$C (ref.\cite{kim:20b}) or 850$^{\circ}$C (ref.\cite{song:23}) at 400$^{\circ}$C/min$^{-1}$. Processes monitored by RHEED\footnote{RHEED: Reflection high-energy electron diffration.}, alternatively repeated until desired film thickness.  $\bullet$ Relaxed films. SPE: PLD at $T\leq 600^{\circ}$C, followed by annealing at $T\geq 800^{\circ}$C in a sealed tube with IrO$_2$ powder in air.\cite{kim:20b} \\

&$\bullet$ SPE: PLD (550$^{\circ}$C, 50 mTorr, 1.1 Jcm$^{-2}$, 5 Hz), post-annealing at 1000$^{\circ}$C, 1.5 h in air.\cite{ghosh:23a}\\ \cmidrule[0.5pt]{1-2}

\multirow{2}{*}{Eu$_2$Ir$_2$O$_7$ | YSZ(111)} &$\bullet$ SPE: PLD (Ir-rich target: Eu:Ir=1:3, 500$^{\circ}$C, 100 mTorr Ar gas containing 1\% 0$_2$, 6 Jcm$^{-2}$ and 10 Hz), post-annealing at 1000$^{\circ}$C, 1.5 h in air.\cite{fujita:15,fujita:16b}\\ 
&$\bullet$ SPE:\cite{liu:21}  same conditions as in Y$_2$Ir$_2$O$_7$ (Ref.\cite{liu:20}).\\
&$\bullet$ SPE: PLD (Ir-rich target: Eu:Ir=1:3, 550$^{\circ}$C, 50 mTorr, 1.1 Jcm$^{-2}$ and 5 Hz), post-annealing at 1000$^{\circ}$C, 1.5 h in air with IrO$_2$ podwer.\cite{ghosh:22a,ghosh:22b,ghosh:23b}\\
&$\bullet$ SPE: PLD, alternate ablation of IrO$_2$ and Eu$_2$O$_3$ targets at subtrate temperature in the range 450-550$^{\circ}$C and oxygen pressure 20-40 mTorr. Post-annealing in air for a few hours with annealing temperature from 800 to 1200$^{\circ}$C.\cite{wu:24}\\
\cmidrule[0.5pt]{1-2}
Tb$_2$Ir$_2$O$_7$ | YSZ(111) &$\bullet$ SPE:\cite{kozuka:17,fujita:18}  same conditions as in Eu$_2$Ir$_2$O$_7$ (Ref.\cite{fujita:15,fujita:16b}).\\\cmidrule[0.5pt]{1-2}

Dy$_2$Ir$_2$O$_7$ | YSZ(111) &$\bullet$ SPE: PLD, target with precursors at a ratio of 3:1=Ir:Dy, laser fluence of 1.5 Jcm$^{-2}$, 3 Hz, substrate temperature fixed at 650$^{\circ}$C, and oxygen pressure of 0.05 mbar during the deposition.  Post-annealing {\text{ex-situ}} at 1000$^{\circ}$C for 1.5 h.\\\cmidrule[0.5pt]{1-2}

Y$_2$Ir$_2$O$_7$ | YSZ(111) &$\bullet$ SPE: PLD (Ir-rich target: Y:Ir=1:3, 480-520$^{\circ}$C, 50-100 mTorr Ar:0$_2$=10:1, 6 Jcm$^{-2}$ and 10 Hz), post-annealing at 1000$^{\circ}$C for 2 h in air or at 900–950$^{\circ}$C for 8–10 min under 650 Torr atmosphere of O$_2$ inside the deposition chamber.\cite{liu:20}\\
\bottomrule
\end{tabular}
\end{table*}

\subsection{\label{subsec:growth_films}Epitaxial growth}

Translating the physical properties of bulk materials into high-quality epitaxial films presents significant challenges due to the formation of intrinsic defects.  These include point defects, which accommodate structural imperfections and  non-stoichiometry, and strain-relieving misfit dislocations, both of which can degrade film properties.\cite{oka:17,hensling:24}  The level of  difficulty varies for each system, but overcoming these challenges is highly rewarding, as dimensional confinement and interface effects can enhance or induce certain material properties that are not observed in bulk materials.

\begin{figure}[htb]
 \includegraphics[keepaspectratio=true, width=0.83\linewidth]{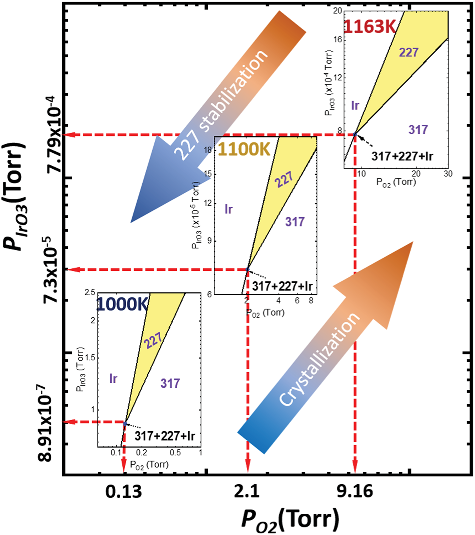} \caption{\label{Fig_2021_Guo_fig_5}  {{\bf{Calculated phse diagram of the system Pr-IrO$_3$-O$_2$ at 1000, 1100 and 1163 K.}} According to calculations, partial pressure P$_{\mbox{IrO}_3}$ for the formation of Pr$_2$Ir$_2$O$_7$ decreases three orders of magnitude and P$_{\mbox{O}_2}$ decreases by 70 times when temperature goes from 1163 K down to 1000 K. However, films deposited {\textit{in situ}} at temperatures lower than 1073 K exhibited low quality.   L. Guo {\textit{et al.}}, npj Computational Materials, 7(1), 10.1038/s41524-021-00610-9, 2021; licensed under a Creative Commons Attribution (CC BY) license.\cite{guo:21}
}}
\end{figure}

Common thin film synthesis approaches for growing crystalline oxide films\cite{schlom:08,martin:10,chambers:10} include pulsed laser deposition (PLD),\cite{lowndes:96,christen:08,groenen:15} sputtering,\cite{brewer:17} and molecular beam epitaxy (MBE).\cite{schlom:15,brahlek:18,rimal:24}  In these physical vapour deposition (PVD) methods, the oxygen partial pressure and substrate temperature define the thermodynamic window for the growth of different materials and are crucial in determining the phases that form in the film.

Growing high-quality single crystalline films of pyrochlore iridates needs both high temperatures and high oxygen partial pressure, as iridium is hard to oxidize. At these elevated temperatures, iridium metal, known for its highly volatility in oxygen, forms the volatile oxide IrO$_3$.\cite{alcock:60,cordfunke:62}  This high vapour pressure of IrO$_3$ is the primary challenge related to Ir oxidation:\cite{wimber:74} IrO$_3$ gas is pumped away from the vacuum chamber, resulting in the depletion of Ir in the deposited films. Consequently, \textit{in situ} growth of pyrochlore iridates by PLD\cite{fujita:15,ohtsuki:20} and sputtering\cite{guo:21} have been hindered.

For instance, PLD carried out to grow Eu$_2$Ir$_2$O$_7$ films at temperatures in the range 1000-1250$^{\circ}$C and oxygen partial pressure between $10^{-7}-10^{-1}$ Torr results in the crystallisation of metallic Ir and polycrystalline Eu$_2$O$_3$.  Increasing the pressure up to 1 Torr at the same temperature range leads to the evaporation of Ir and the formation of crystalline Eu$_2$O$_3$ only.\cite{fujita:15}  Additionally, using an Ir-rich target\cite{ohtsuki:20} or a separate IrO$_2$ target\cite{guo:21} does not overcome these difficulties for \textit{in situ} PLD growth of pyrochlore iridates.

Perovskite-related phases of iridates, such as SrIrO$_3$ and Sr$_2$IrO$_4$, along with the binary oxide IrO$_2$, have been successfully grown using MBE.\cite{kawasaki:16} Recently, this method was also employed to grow a pyrochlore titanate.\cite{anderson:24} However, MBE has yet to be reported for the growth of pyrochlore iridates.

Guo \textit{et al.}\cite{guo:21} carried out computational calculations and experimental growths in the Pr-Ir-O$_2$ system by simultaneous sputtering from both Pr$_2$Ir$_2$O$_7$ and IrO$_2$ targets. They also found that \textit{in-situ} synthesis of Pr$_2$Ir$_2$O$_7$ at high temperature (1163 K) needs oxygen partial pressures $\geq 9$ Torr, making it unfeasible in PVD systems, Fig.\ref{Fig_2021_Guo_fig_5}. Pr$_2$Ir$_2$O$_7$ can only form if both partial pressures of gas species O$_2$ and IrO$_3$(g) are high, otherwise Ir(s) forms at low oxygen partial pressures. Moreover, lower growth temperature (1073 K) , which  would allow for lower partial pressures, resulted in films with poor crystallinity.

Kim \textit{et al.}\cite{kim:19} studied the chemical kinetic instabilities of IrO$_2$ thin films, by \textit{in-operando} spectroscopic ellipsometry.  They found that conventional \textit{in-situ} growth techniques are inadequate for growing pyrochlore iridate thin films.  This is due to the unavoidable dissociation of IrO$_2$ and formation of IrO$_3$(g) at high temperatures, while crystallization of the film is not possible at lower temperatures.

\begin{figure}[htb]
 \includegraphics[keepaspectratio=true, width=\linewidth]{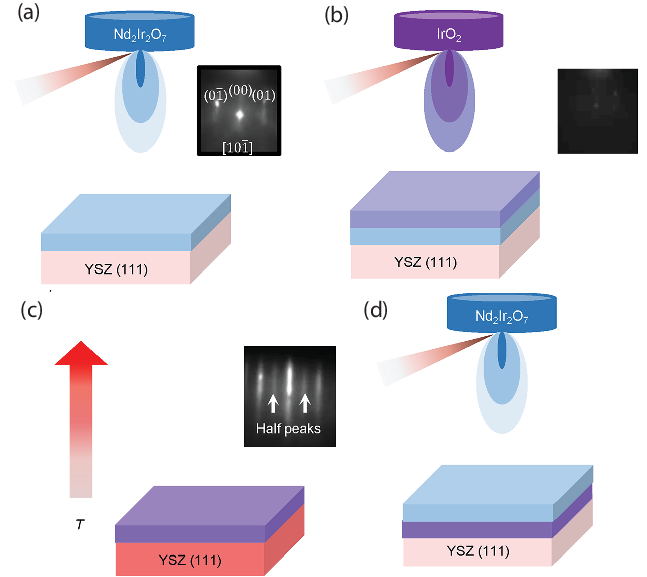} \caption{\label{Fig_2023_Song_fig_1}  {{\bf{Steps of the Repeated Rapid High Temperature Synthesis Epitaxy (RRHSE) method for growing Nd$_2$Ir$_2$O$_7$ thin films.}} (a) Conventional PLD of a Nd$_2$Ir$_2$O$_7$ target at 600$^{\circ}$ C and oxygen partial pressure of 50 mTorr. (b) Conventional PLD of an IrO$_2$ target to compensate for Ir loss. (c) Rapid thermal annealing for approximately $30$ s, heating up to 850$^{\circ}$ C at a rate of 400 $^{\circ}\;\mbox{C/min}$ . (d) Repeat steps (a) to (c) steps until the derised thickness of the film is achieved.    J. K. Song {\textit{et al.}}, APL Materials, 11(6), 10.1063/5.0153164, 2023; licensed under a Creative Commons Attribution (CC BY) license.\cite{song:23}
}}
\end{figure}

To address this challenges of epitaxial growth techniques for pyrochlore iridates, the synthesis is divided into two steps using the solid-state epitaxy method.  First, a layer with the correct Ir stoichiometry is formed, followed by its crystallisation.\cite{hellman:96,evans:18,kim:22}  In this approach, an amorphous layer, grown at a much lower temperature than needed for epitaxial growth, transforms into an epitaxial thin film through a high temperature post-annealing process, with the single crystal substrate serving as a seed. The amorphous thin films, which serve as precursors for crystallization, are formed by PLD or sputtering.

An alternative approach is called repeated rapid high-temperature synthesis epitaxy (RRHSE, Fig. \ref{Fig_2023_Song_fig_1}).\cite{kim:20b,song:23,shen:24}  This method involves a rapid thermal annealing, performed {\textit{in situ}}. The amorphous layer formed from the PLD previous step at low temperature (aproximately $600^{\circ}$ C) is heated up to $850^{\circ}$ C at a rate of 400 $^{\circ}\;\mbox{C/min}$ for about $30$ s using an infrared laser heater in the deposition chamber. Then the temperature is then reduced to $600^{\circ}$ C for a new cycle of ablation followed by a rapid thermal annealing. In this synthesis process, as the annealing is carried out under pumping of IrO$_3$(g), the Ir loss is compensated by ablating an IrO$_2$ target before the annealing step. In contrast to the standard solid-state epitaxy approach, which results in relaxed films, RRHSE aenables the growth of strained films.

Table \ref{tab:table1} provides a detailed summary of the epitaxial pyrochlore iridate films synthesized to date.

Another critical issue in synthesising pyrochlore iridate films is that even minute variations in the ratio of the rare-earth element to iridium can significantly impact their electronic and magnetic properties. Cation disorder, in which an excess of rare-earth ions occupies the Ir sites in R$_{2+x}$Ir$_{2-x}$O$_{7-x/2}$, commonly referred to as \textit{stuffing}, can bring about substantial changes in the ground state properties of these materials.  For instance, the ground state of Pr$_2$Ir$_2$O$_7$ is still under discussion, due to the difficulty in achieving the correct  stoichiometry, even in polycrystalline samples.\cite{macLaughlin:15}  Experimental results for polycrystalline Eu$_2$Ir$_2$O$_7$ were also found to depend on the synthesis parameters. Specifically, deviations from the ideal Eu/Ir ratio increased the variability in its metallic behaviour above the metal-to-insulator transition.\cite{telang:18}  Moreover, the ordering temperature of a single crystal of Tb$_2$Ir$_2$O$_7$ decreased from 130 K in the stoichiometric material to $\sim$71 K with an excess of Tb of 0.18, although it retained the AIAO magnetic structure.\cite{donnerer:19}  Strong sample dependence of the magnetoransport properties was also found in single crystals of Pr$_2$Ir$_2$O$_7$. Minor antisite mixing between Pr and Ir resulted in local structural imperfections that influence the Ir 5d topological electronic band, which is modulated by the interaction between Ir conduction electrons and localized Pr 4f moments. Therefore, a precise structural analysis was conducted to select a near-stoichiometric sample.\cite{ueda:22a}

\subsection{\label{film_properties}Experimental findings in epitaxial pyrochlore iridate films}

The investigation of fundamental effects in pyrochlore iridates in thin film form raises many interesting questions and is still in its early stages. Current research has primarily focused on the magnetic Weyl semimetal state and its associated topological phenomena, such as the anomalous Hall effect and negative magnetoresistance that can provide evidence for the chiral anomaly (charge transfer between Weyl fermions with opposite chirality).

\begin{figure}[htb]
 \includegraphics[keepaspectratio=true, width=0.6\linewidth]{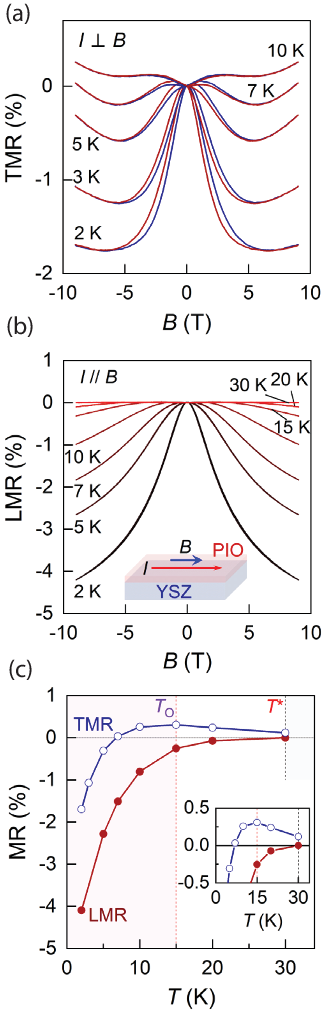} \caption{\label{Fig_2021_Li_Correlated_fig}  {{\bf{Magnetotrasnport in strained Pr$_2$Ir$_2$O$_7$ films.}} (a) Transverse magnetoresistance (TMR) at different temperatures. Blue (red) lines correspond to field decreasing (increasing) magnetic field sweeps. (b) Longitudinal magnetoresistance (LMR) as a function of temperature. (c) TMR and LMR measured as 9 T as a function of temperature.  Y. Li {\textit{et al.}}, Advanced Materials, 33(25), 10.1002/adma.202008528, 2021; licensed under a Creative Commons Attribution (CC BY) license.\cite{li:21} 
}}
\end{figure}

Pr$_2$Ir$_2$O$_7$ thin films, grown via solid-state epitaxy, show topological nontrivial transport properties.\cite{ohtsuki:19}  Certain regions of the Pr$_2$Ir$_2$O$_7$ film are stained to the substrate, as indicated by scanning transmission electron microscopy, breaking the cubic symmetry due to epitaxial compressive strain in the in-plane direction (tensile strain along the surface normal [111] direction), and exhibiting spontaneous Hall effect under zero magnetic field.  The Hall resistivity increases with decreasing temperature from 50 K down to 700 mK without any observable spontaneous magnetization. This observation is consistent with the emergence of the magnetic Weyl semimetal phase.  The authors attributed the breaking of time-reversal symmetry, necessary for the appearance of a Weyl semimetal phase, to the magnetic order of Ir 5d electrons, although this point is unsettled.\cite{guo:20} Additionally, the Weyl semimetal phase can be induced by a magnetic field collinear to the electric current, giving rise to negative longitudinal magnetoresistance, arising from the chiral anomaly.\cite{nagaosa:20}

\begin{figure}[htb]
 \includegraphics[keepaspectratio=true, width=\linewidth]{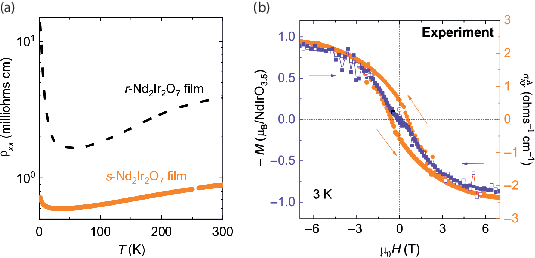} \caption{\label{Fig_2020_Kim_fig}  {{\bf{Transport properties of strained Nd$_2$Ir$_2$O$_7$ thin films.}} (a) Longitudinal resistivity $\rho_{xx}$ as a function of temperature for a fully strained Nd$_2$Ir$_2$O$_7$ film (orange). $\rho_{xx}$ for a relaxed film is also shown (dashed, black). (b) Magnetization ($M$, blue) and anomalous Hall conductivity ($\sigma_{xy}^{A}$, orange) of a strained film with the current along $[\bar{1}10]$ and magnetic field applied along $[111]$. The anomalous Hall conductivity was calculated as $\sigma_{xy}^{A}=\rho_{xy}^{A}/((\rho_{xx})^2 +(\rho_{xy}^{A})^2)$, where the anomalous Hall resistivity $\rho_{xy}^{A}$ is derived from the measured Hall resistivity $\rho_{xy}$ after antisymmetrizing using positive and negative field sweep branches and removing the linear part, corresponding to the conventional Hall resistivity.   W. J. Kim {\textit{et al.}}, Science Advances, 7(29), 10.1126/sciadv.abb1539, 2020; licensed under a Creative Commons Attribution (CC BY) license.\cite{kim:20b} 
}}
\end{figure}

Guo {\textit{et al.}} observed an increase in the onset temperature of the spontaneous Hall effect by one order of magnitude in epitaxial films of Pr$_2$Ir$_2$O$_7$ compared to the bulk material, from 1.5 K to 15 K.\cite{guo:20} This effect likely arises from the topological Hall effect, in which time-reversal symmetry is broken from the frustrated spin-liquid correlations rather than a net magnetic moment. A minute lattice distortion of the Ir sublattice, caused by epitaxial strain, localizes the Ir moments although no sign of long-range AIAO order is observed. Thus, two hypotheses are proposed as the origin of the time-reversal symmetry breaking in these films: Ir-site spin correlations or a modification of the Pr-Pr coupling through spin polarization at the Ir site. Further investigation of the magnetism in these epitaxial films is needed.

While AIAO magnetic order is not observed in unstrained bulk or single crystals of Pr$_2$Ir$_2$O$_7$, mean-field calculations for Pr$_2$Ir$_2$O$_7$ under biaxial strain, based on the Hubbard model predict a strain-induced AIAO antiferromagnetic order in Pr$_2$Ir$_2$O$_7$ films with small strain ($< 0.3$\%) that breaks time-reversal symmetry, leading to a WSM phase.\cite{li:21}  To test this prediction, films with a strain of $\approx 0.2$ \% were synthesised. These films exhibit negative transverse magnetoresistance (TMR) with a hysteresis feature below 15 K when magnetic field is applied along the [111] direction (Fig. \ref{Fig_2021_Li_Correlated_fig}(a)), contrasting with the positive TMR observed in bulk Pr$_2$Ir$_2$O$_7$.  The contribution of the AIAO magnetic ordering is dominant below 7 K, with coexistence of paramagnetic and AIAO phases in the range of 7–15 K.  The hysteresis behaviour is consistent with the presence of two degenerate magnetic domains in the AIAO antiferromagnetic state, although the AIAO ordering was not confirmed by other techniques. Negative longitudinal magnetoresistance without hysteresis is observed below 30 K (Fig. \ref{Fig_2021_Li_Correlated_fig}(b)), which is attributed to the chiral anomaly of Weyl fermions in the WSM phase, once the influence of current jetting on longitudinal resistivity measurements is ruled out.

Empirical evidence supporting topological phenomena has also been observed in antiferromagnetic Nd$_2$Ir$_2$O$_7$ films.\cite{kim:20b} A significantly larger anomalous Hall conductivity was found in compressively strained Nd$_2$Ir$_2$O$_7$ films compared to relaxed films, suggesting that biaxial strain affects the Berry curvature. The AIAO spin structure of the Ir sublattice is modulated under stain, leading to the formation of magnetic multipoles that can induce anomalous Hall effect (AHE) without magnetization. The authors applied the theoretical framework of atomic cluster multipoles in antiferromagnetic crystals, as proposed by Suzuki {\textit{et al.}},\cite{suzuki:17} to the noncollinear magnetic structure of Nd$_2$Ir$_2$O$_7$ films. It is suggested that octupoles moments with the same symmetry as magnetic dipole moments can induce AHE without magnetization. These multipoles form under strain but not in relaxed films, which exhibit negligible spontaneous Hall conductivity.

More recently, a linear scaling between saturated anomalous Hall conductivity at 9 T and Hall carrier density was found in strained Nd$_2$Ir$_2$O$_7$.\cite{shen:24} This linear slope changes its sign around the ordering temperature of Nd moments. This humplike feature was attributed to a change of the energy levels of Weyl nodes, according to numerical calculations under epitaxial strain that breaks the cubic symmetry.  

The conductivity of domain walls (DW) in Nd$_2$Ir$_2$O$_7$ single crystals has been experimentally found to be one order of magnitude higher than in the bulk material.\cite{ueda:15b,tian:16} DW-induced magnetotransport in Nd$_2$Ir$_2$O$_7$ films has also been investigated.\cite{kim:18} In these films, AHE from both bulk and DW has been reported under an applied magnetic field along the [111] direction, which is expected to favour the formation of DW within the (111) plane. It is proposed that hysteretic, humplike behaviour of AHE observed in these films originates from DW conduction.

\begin{figure}[htb]
 \includegraphics[keepaspectratio=true, width=1.0\linewidth]{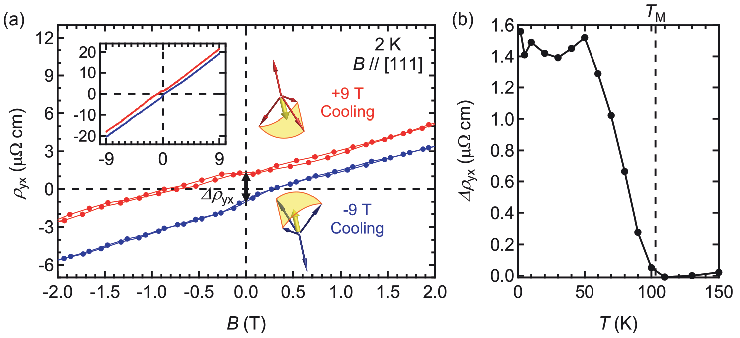} \caption{\label{Fig_Fujita_odd_fig_4}  {{\bf{Hall resistivity in Eu$_2$Ir$_2$O$_7$ single crystalline thin films.}} (a) Field cooling at $+9$ T or $-9$ T is applied, giving rise to A or B domains of the AIAO spin structure. The Hall resistance exhibits a non-zero value at 0 T, $\Delta \rho_{xy}$. (b) Temperature dependence of $\Delta \rho_{xy}$ (dashed line at the transition temperature).  T. C. Fujita {\textit{et al.}}, Scientific Reports, 5, 10.1038/srep09711, 2015; licensed under a Creative Commons Attribution (CC BY) license.\cite{fujita:15}
}}
\end{figure}

In epitaxial films of Eu$_2$Ir$_2$O$_7$, it was found that the magnetic domain structure (AIAO or AOAI) can be selectively stabilized by the polarity of the cooling field along the [111] direction above $\pm3$ T (a multidomain structure is formed by cooling in lower magnetic fields).\cite{fujita:15} The domain structure formed in this way remains robust against a magnetic field of at least 9 T below the temperature at which the metal-to-insulator transition occurs.  These robustness in Eu$_2$Ir$_2$O$_7$ has been attributed to the nonmagnetic character of Eu$^{3+}$.  As conduction electrons are sensitive to the local spin structure due to spin-orbit coupling, the magnetic domain structure was detected by magnetotransport measurements. The magnetoresistance contains a symmetric term independent of the cooling field and a linear term whose sign changes by inverting the polarity of a cooling field above $\pm3$ T.  The Hall resistance also exhibits a non-zero value at 0 T which depends on lattice distortion due to epitaxial strain, leading to a scalar spin chirality that does not cancel out. Additionally, AHE with no net magnetization has also been reported in other works on Eu$_2$Ir$_2$O$_7$ films.\cite{liu:21}

Fujita {\textit{et al.}}\cite{fujita:18} investigated the magnetic domain structure through the coupling between Tb$^{3+}$ and Ir$^{4+}$ moments in Tb$_2$Ir$_2$O$_7$ single-crystalline thin films, in which both sublattices form the AIAO magnetic structure at temperatures around 120 and 40 K, respectively, similar to the bulk material.  Specifically, they studied how molecular fields of Tb$^{3+}$ moments affect the magnetoresponse of Ir$^{4+}$ below the ordering temperature of Tb$^{3+}$, which exhibits axial magnetocrystalline anisotropy along the [111] direction. The magnetic domain structure of the Ir$^{4+}$ sublattice was controlled through the cooling procedure and analysed by the sign of the linear term of magnetoresistance.\cite{arima:13,fujita:15} It was observed that Ir$^{4+}$ moments experience opposite fields at temperatures lower or higher that the ordering temperature of Tb$^{3+}$.  These results suggest that molecular fields from Tb$^{3+}$ are antiparallel to the external field along the [111] direction. Additionally, in contrast to  Nd$_2$Ir$_2$O$_7$, little contribution from domain-wall conduction was observed in Tb$_2$Ir$_2$O$_7$.

\section{Future research directions}

\begin{table*}
\caption{\label{table_doped}Bulk polycrystalline hole-doped pyrochlore iridates}
\renewcommand{\arraystretch}{1.3}
\begin{tabular}{p{4cm}p{12cm}}
\toprule
Material &  \thead{Key experimental evidences} \\\midrule
(Nd$_{1-x}$Pr$_x$)$_2$Ir$_2$O$_7$ & Seebeck and Nernst signals are amplified for $x=0.5$  in proximity to the field-induced metal-to-insulator transition.\cite{ueda:22b}  \\
\cmidrule[0.5pt]{1-2}
(Eu$_{1-x}$Ca$_x$)$_2$Ir$_2$O$_7$, (Tb$_{1-x}$Ca$_x$)$_2$Ir$_2$O$_7$, (Gd$_{1-x}$Cd$_x$)$_2$Ir$_2$O$_7$ & (Gd$_{1-x}$Cd$_x$)$_2$Ir$_2$O$_7$ and (Tb$_{1-x}$Ca$_x$)$_2$Ir$_2$O$_7$ exhibit a systematic reduction in resistivity and in the magnetic transition temperature as $x$ increases and turn into a paramagnetic phase at sufficiently large $x$. For $x= 0.12$, (Gd$_{1-x}$Cd$_x$)$_2$Ir$_2$O$_7$ displays a large Hall effect driven by the magnetic exchange coupling between Gd 4$f$ and Ir 5$d$ moments.\cite{ueda:20}
  \\
\cmidrule[0.5pt]{1-2}
(Eu$_{1-x}$Ca$_x$)$_2$Ir$_2$O$_7$, ([(Nd$_{0.2}$Pr$_{0.8}$)$_{1-x}$Ca$_x$]$_2$Ir$_2$O$_7$, (Pr$_{1-x}$Ca$_x$)$_2$Ir$_2$O$_7$ & Magnetization and magnetic transition temperature decrease with increasing $x$ in (Eu$_{1-x}$Ca$_x$)$_2$Ir$_2$O$_7$, and the antiferromagnetic transition appears to vanish around $x = 0.05$.  For $x \geq 0.05$, it shows metallic behaviour down to 2 K.  The temperature dependence of the Seebeck coefficient shows a peak consistent with quadratic band touching point in a wide doping range in the doping-induced metallic phase.\cite{kaneko:19} \\ 
\cmidrule[0.5pt]{1-2}

(Eu$_{1-x}$Ca$_x$)$_2$Ir$_2$O$_7$ & MIT is fully suppressed within the doping range $0.04 <x< 0.07$, and the resistivity shows metallic behaviour for $x = 0.07$ and 0.13, with no upturn observed down to 2 K. Short-range antiferromagnetic order (decoupled from the metal-to-insulator transition) persists into the metallic range.\cite{zoghlin:21}\\
\cmidrule[0.5pt]{1-2}
(Eu$_{1-x}$Bi$_x$)$_2$Ir$_2$O$_7$ & The ground state remains insulating for $x<0.035$, while a metallic behaviour appears for $0.1 < x< 1$, with no signs of magnetic ordering. The temperature dependence resistivity changes from $T$ linear ($x = 0.1$) to $T^{3/2}$ ($x \geq 0.5$) with a Fermi-liquid-like $T^2$ dependence at $x = 0.25$.\cite{telang:19,telang:22}  \\
\cmidrule[0.5pt]{1-2}
(Nd$_{1-x}$Pr$_x$)$_2$Ir$_2$O$_7$, (Sm$_{1-x}$Nd$_x$)$_2$Ir$_2$O$_7$ &  Large magnetoresistance is observed for (Nd$_{1-x}$Pr$_x$)$_2$Ir$_2$O$_7$ with $0.4<x<0.7$ near the zero-field phase boundary between the antiferromagnetic insulator and the paramagnetic semimetal. The insulating phase below the transition temperature is systematically suppressed with increasing $x$, and a metallic state with no resistivity upturn down to 2 K and no magnetic order appears around $x=0.8$.  (Sm$_{1-x}$Nd$_x$)$_2$Ir$_2$O$_7$ exhibits a paramagnetic insulator-metal crossover for $y=0.7-0.9$, reminiscent of a first-order Mott transition.\cite{ueda:15a}\\
\cmidrule[0.5pt]{1-2}
(Nd$_{1-x}$Ca$_x$)$_2$Ir$_2$O$_7$ & Suppression of MIT and AIAO order on the Ir sublattice with hole doping. For $0<x<0.08$, as doping level increases, the metal-to-insulator transition temperature decreases. For  $x \geq 0.08$, the ground state becomes metallic with a weak upturn (likely due to disorder) in the low temperature range and signatures of Ir magnetic order vanish.\cite{porter:19}\\
\cmidrule[0.5pt]{1-2}
(Eu$_{1-x}$Sr$_x$)$_2$Ir$_2$O$_7$& Metal-insulator transition temperature reduced with hole doping. Non-Fermi liquid behaviour for x = 0.2.\cite{banerjee:17} \\
\cmidrule[0.5pt]{1-2}
(Eu$_{1-x}$Nd$_x$)$_2$Ir$_2$O$_7$& Metal-insulator transition temperature decreases with increase in Nd content. Linear specific heat in the insulating region.\cite{mondal:20} \\
\bottomrule
\end{tabular}
\end{table*}

\subsection{Chemical doping}

Carrier doping has the potential to produce metals with emergent properties, enabling control of the band filling in the Ir 5d state through the chemical substitution of R$^{3+}$ ions with divalent A ions, such as Ca, in doped pyrochlores (R$_{1-x}$A$_x$)$_2$Ir$_2$O$_7$.  Intermediate band filling-controlled compounds exhibit novel phase transitions and
related magneto-electronic phenomena,\cite{ueda:20} although this has not been yet reported in thin films.  Consequently, the effects of disorder introduced by doping on physical properties remain unexplored. Theoretical studies have also investigated possible phases on a pyrochlore lattice upon doping.\cite{berke:18}  Table \ref{table_doped} summarizes key experimental evidence found in polycrystalline hole-doped pyrochlore iridates, which can guide the exploration of epitaxial doped pyrochlore iridate films.

\subsection{Design and growth of epitaxial heterostructures}

Research on heterointerfaces between pyrochlore oxides holds significant potential for groundbreaking discoveries,\cite{uchida:18,boschker:17} much like those found at interfaces between the extensively investigated family of perovskite oxides.\cite{zubko:11,hwang:12,chakhalian:14,sulpizio:14,hellman:17,huang:18,bhattacharya:14,jeong:23}  But if studies on epitaxial films of pyrochlore oxides are at a very early stage, even more so on their interfaces, since additional challenges arise when integrating these films into heterostructures.  Thus, experimental insight is currently lacking.
b
Theoretically, it was proposed that a metallic surface with topological nature and a ferromagnetic moment, distinct from the bulk antiferromagnetic insulator, will emerge at magnetic domain walls of R$_2$Ir$_2$O$_7$.\cite{yamaji:14} This is triggered by the formation of Fermi arcs at the domain walls, which can be tuned through applied magnetic fields. These domain-wall states, induced by the chiral anomaly, enable Anderson localization by impurities.  Under external magnetic fields they will show anomalous Hall conductivities. 

This magnetic control of the interface electronic transport was studied experimentally in the epitaxial heterostructure Eu$_2$Ir$_2$O$_7$/Tb$_2$Ir$_2$O$_7$.\cite{fujita:16b} In this case, Eu$_2$Ir$_2$O$_7$ acts as the pinned layer and their magnetic domains AIAO or AOAI are determined by the cooling magnetic field, while those of Tb$_2$Ir$_2$O$_7$ are switched by a sweeping magnetic field. The authors reported the observation of additional conduction coming from the interface between domains of different type at both sides of the interface, that was not observed in single layers of neither of these pyrochlores.

This is the only heterostructure between epitaxial pyrochlore iridates experimentally studied so far.

From the theoretical side, formation of a two-dimensional magnetic monopole gas confined at the interface between a pyrochlore spin ice and an antiferromagnetic pyrochlore iridate has been suggested by means of Monte Carlo simulations with 2-in-2-out and all-in-all-out states at each side of the pyrochlore heterointerface.\cite{miao:20}

Epitaxial heterostructures are promising platforms for investigating  interactions between localized frustrated spins in quantum magnets, most of which are insulators, and conducting charge carriers. Therefore, these heterostructures can bridge the gap between fundamental research on insulating quantum magnets and potential electronic applications.

Recently, the interaction between localized moments and itinerant carriers at both sides of an epitaxial heterostructure has been studied as a means to achieve electrical detection of spin states in insulating quantum magnets. Zhang {\textit{et al.}}\cite{zhang:23} investigated an epitaxial pyrochlore structure in which ultra thin films of nonmagnetic metallic Bi$_2$Ir$_2$O$_7$ were grown on Dy$_2$Ti$_2$O$_7$ single crystals that exhibit spin ice behaviour. They observed anomalies in the magnetoresistance of Bi$_2$Ir$_2$O$_7$ when a magnetic field was applied along the [111] direction, at the same field intensity at which anomalies were observed in the AC susceptibility. These findings suggest that charge carriers from the metallic pyrochlore are sensitive to the spin flips of the spin ice pyrochlore, transitioning from the 2-in-2-out configuration to the 3-1/1-3 state.

\begin{figure}[htb]
 \includegraphics[keepaspectratio=true, width=1.0\linewidth]{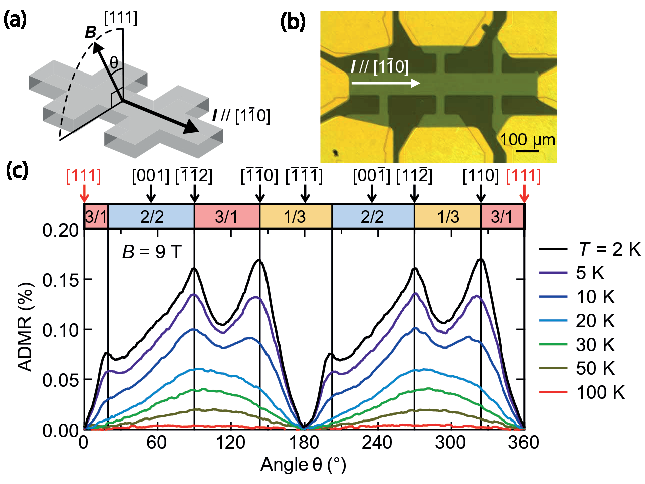} \caption{\label{2024_Ohno_Proximity_effect}  {{\bf{Proximity efffect at the epitaxial pyrochlore heterostructure Bi$_2$Rh$_2$O$_7$/Dy$_2$Ti$_2$O$_7$.}} (a) Measurement configuration of the field-angle dependent resistivity performed under a magnetic field.  (b) Hall device structure used in the experiment.  (c) Angle dependent magnetoresistance ($\mbox{ADMR}= \rho{xx}(\theta)/\rho_{xx}(0^{\circ}) -1$)  for several temperatures at 9 T.
Crystallographic directions and expected magnetic structures of the Dy 4$f$ moments are indicated in the upper part of the plot. 
M. Ohno {\textit{et al.}}, Science Advances, 10(11), 10.1126/sciadv.adk6308, 2024; licensed under a Creative Commons Attribution (CC BY-NC) license.\cite{ohno:24}
}}
\end{figure}

Ohno {\textit{et al.}}\cite{ohno:24} also explored the electrical detection of magnetic transitions in the spin ice insulator Dy$_2$Ti$_2$O$_7$ through its interface with a nonmagnetic pyrochlore metal. In this study, a relaxed Dy$_2$Ti$_2$O$_7$ film was grown on a La$_2$Zr$_2$O$_7$ buffer layer on a YSZ (111) substrate, ensuring no epitaxial strain occurs and allowing Dy$_2$Ti$_2$O$_7$ to exhibit magnetic properties similar to those of the bulk material.  Subsequently, a film of Bi$_2$Rh$_2$O$_7$ (hosting a 4$d$ metal at the $B$ site instead of Ir) was epitaxially grown on Dy$_2$Ti$_2$O$_7$.  A similar heterostructure could be made with a pyrochlore iridate, and this example is noteworthy as it demonstrates the functionalization of a pyrochlore quantum magnet through interface transport phenomena.   

Specifically, changes of the spin structure in Dy$_2$Ti$_2$O$_7$ were controlled by a magnetic field along the [111] direction, and its magnetic transitions were detected through the topological Hall effect in Bi$_2$Rh$_2$O$_7$, thanks to the proximity effect at the interface Bi$_2$Rh$_2$O$_7$/Dy$_2$Ti$_2$O$_7$.  The noncoplanar spin configuration gives rise to an internal effective magnetic field, proportional to the spin chirality (solid angle subtended by the spins), whose sign changes when a magnetic field is applied along [111], but not when the field is along the [001] direction.  Additionally, field-angle dependent longitudinal resistivity measured at a field of 9 T shows peaks that can be attributed to the spin transitions in Dy$_2$Ti$_2$O$_7$ from the 2-in/2-out (2/2) to the 3-in/1-out (3/1) configuration (Fig.\ref{2024_Ohno_Proximity_effect}), in contrast to the isotropic transport observed at the interface of Bi$_2$Rh$_2$O$_7$ with the nonmagnetic Eu$_2$Ti$_2$O$_7$.  This study bridges fundamental research on insulating quantum magnets with potential electronic applications.  The findings could lead to transformative innovations in quantum technologies by enabling electrical detection of emergent quantum phenomena.

\subsection{Substrates for straining pyrochlore iridates}

Epitaxial strain is a crucial parameter for inducing topological phenomena in pyrochlore iridates, as structural distortions lead to changes in the electronic structure due to electron-lattice coupling and break the cubic symmetry.  Therefore, the growth of single-crystal pyrochlore substrates is an essential enabling technology for thin film investigations, facilitating the exploration of new phases.

Recently, pyrochlore substrates of Y$_2$Ti$_2$O$_7$ have become commercially available. Additionally, other pyrochlore oxides, such as Tb$_2$Ti$_2$O$_7$, Gd$_2$Ti$_2$O$_7$, Dy$_2$Ti$_2$O$_7$ and Ho$_2$Ti$_2$O$_7$,\cite{klimm:17,guo:14,kang:14a,kang:14b} have been grown as bulk crystals using the Czochralski method. Thus, it is expected that single crystal substrates with pyrochlore structure will become more common in the near future, enabling growth thin films with tunable strain.

Alternatively, buffering a fluorite substrate with a pyrochlore oxide can also achieve different levels of structural distortions in the epitaxial films.\cite{ohno:23} Although this approach may compromise the functional of a heterostructure due to interfacial roughness, it has been validated as an effective method for providing an intermediate layer in the growth of a pyrochlore spin ice.\cite{ohno:24} In that case, the buffer layer was employed to obtain a pyrochlore spin ice film free from epitaxial strain with magnetic properties similar to those of bulk single crystals.

\section{Conclusions}

Research on pyrochlore iridates can advance by building upon previous successful approaches that have been applied to simpler materials, such as studing novel phenomena at heterointerfaces.  The outcomes of this research will indicate new research directions for functional materials, since interfacing oxides with pyrochlore structure remains an unexplored area.  Identifying combinations of materials with desired properties for specific applications will mark a breakthrough. 

On the other hand, the technological relevance of pyrochlore oxides stems from the unusual types of fractionalized quasiparticles (emergent excitations of collective behaviour) that they can support, such as magnetic monopoles in pyrochlore titanates with a spin ice state or Weyl fermions in pyrochlore iridates with a magnetic Weyl semimetal state. These materials can potentially drive major technological innovations. It is hoped that this review will stimulate further discussions and research in this field.

\section*{Author declarations}
\subsection*{Conflict of Interest}

The authors have no conflicts to disclose

\section*{Data availability}

Data sharing is not applicable to this article as no new data
were created or analyzed in this study.

\bibliography{biblio_pyro_v_01}

\end{document}